\documentclass[]{agujournal2019}
\usepackage{url} 
\usepackage{lineno}
\usepackage{amsmath}
\usepackage[finalnew]{trackchanges}
\usepackage{soul}
\usepackage{ragged2e}

\draftfalse

\journalname{JGR Planets}

\begin{document}

\title{A Recharge Oscillator Model for Interannual Variability in Venus' Clouds}

\authors{Pushkar Kopparla\affil{1,2}, Ashwin Seshadri\affil{3}, Takeshi Imamura\affil{1} and Yeon Joo Lee\affil{4}}

 \affiliation{1}{Graduate School of Frontier Sciences, The University of Tokyo, Kashiwa, Japan}
 \affiliation{2}{Center for Space and Habitability, University of Bern, Bern, Switzerland}
 \affiliation{3}{Centre for Atmospheric and Oceanic Sciences and Divecha Centre for Climate Change, Indian Institute of Science, Bengaluru, India}
 \affiliation{4}{Zentrum f\"ur Astronomie und Astrophysik,
Technische Universit\"at Berlin, Berlin, Germany}

\correspondingauthor{Pushkar Kopparla}{pushkarkopparla@gmail.com}




\begin{keypoints}
\item A simple model is {developed} to explore relationships between convective activity in the cloud layer and cloud-base water abundance.
\item Sustained recharge-discharge oscillations exist within this model on interannual to decadal timescales.
\item {The interannual to decadal sulfur dioxide variability at the cloud tops could be a result of these oscillations, instead of explosive volcanic injections.}
\end{keypoints}

\begin{abstract}
Sulfur dioxide is a radiatively and chemically important trace gas in the atmosphere of Venus and its abundance at the cloud-tops has been observed to vary on interannual to decadal timescales. This variability is thought to come from changes in the strength of convection which transports sulfur dioxide to the cloud-tops, {although} the dynamics behind such convective variability are unknown. Here we propose a new conceptual model for convective variability that links the radiative effects of water abundance at the cloud-base to convective strength within the clouds, which in turn affects water transport within the cloud. The model consists of two coupled equations which are identified as a recharge-discharge oscillator. The solutions of the coupled equations are finite amplitude sustained oscillations in convective strength and cloud-base water abundance on 3-9 year timescales. The characteristic oscillation timescale is given by the geometric mean of the radiative cooling time and the eddy mixing time near the base of the convective clouds. 
\end{abstract}

\section*{Plain Language Summary}
Water and sulfur dioxide are important trace gases in the atmosphere of Venus. The photolysis of sulfur dioxide in the upper clouds produces sulfuric acid, which forms the thick cloud decks characteristic of the planet's atmosphere. Sulfur dioxide abundances at the cloud-top of Venus (about 70 km altitude) have been observed to oscillate on interannual to decadal timescales. In this paper, we use a simplified model of atmospheric dynamics and chemistry to outline the mechanism that causes such oscillations. The water abundance at the base of the clouds (about 47 km altitude), has a strong influence on the cloud-base heating and cloud level convection. The cloud level convective mixing in turn determines the gradient of water abundance in the cloud layer, and thereby the cloud-base water abundance. Thus, the convection and water abundance form a  coupled system {that} oscillates on interannual to decadal timescales,  which can explain the timescale of variability in the transport of sulfur dioxide to the cloud-tops.

\section{Introduction}
The clouds of Venus are among the primary controls of the atmospheric radiative balance, and are composed of sulfuric acid, water and other sulfur-based aerosols which form from the photolysis of sulfur dioxide \cite{esposito198316}. The main cloud deck can be resolved into three distinct regions, and ranges from 70 to 47 km in the atmosphere \cite{knollenberg1980microphysics}. The upper clouds are formed by the photochemical production of sulphuric acid from sulphur dioxide, the middle clouds by the droplet growth in the convective region and lower clouds by condensation of sulphuric acid from the lower atmosphere on to the middle cloud droplet flux  \cite{krasnopolsky1994h2o,imamura2001microphysics,mills2007atmospheric}.  
Climate modeling studies of Venus have noted that the climate state is very sensitive to perturbations of SO$_2$, H$_2$O and cloud albedo  \cite{hashimoto2001predictions,bullock2001recent}. Several decades of ground and space based observations of sulfur dioxide show that this trace gas is highly variable at the cloud-tops at timescales of hours to decades \cite[and references therein]{encrenaz2020hdo,marcq2020climatology,encrenaz2016hdo,vandaele2017sulfur2,vandaele2017sulfur}, varying by up to two orders of magnitude. Understanding the nature of this variability is critical to understanding the trajectory of Venus climate.

 The deep atmosphere of Venus ($\sim$ 40 $km$) has sulfur dioxide and water abundances with mean values and variabilities of approximately 130$\pm$50 and 30$\pm10$ $ppm$ (parts per million) \cite{marcq2008latitudinal,barstow2012models,marcq2018composition}. These gases are then raised up to cloud-tops at low latitudes by convective uplifting, {where} their concentrations decrease due to photodissociation to form sulfuric acid and other sulfur aerosols. At the cloud-tops, sulfur dioxide abundance varies between 10-1000 $ppb$ \cite{vandaele2017sulfur} and water 1-7 $ppm$ \cite{fedorova2008hdo,encrenaz2016hdo,fedorova2016variations}. The short\add[referee2]{-}term (hourly-daily) variability of sulfur dioxide is likely due to the localized variations of vertical mixing via small convective cells and its fast dissociation by photolysis \cite{marcq2013variations,vandaele2017sulfur2}. Indeed, such small scale convective variability is seen in recent radio observations of the cloud layer \cite{imamura2018fine}. On decadal timescales, sulfur dioxide variability at the cloud-top is characterized by strong intermittent injections from the troposphere followed by slow decays due to photolysis \cite{marcq2013variations,vandaele2017sulfur2}. While early studies attributed such injections to episodic volcanic activity  \cite{esposito1984sulfur}, more recent literature has shifted towards interpreting it in terms of variability of vertical mixing in the cloud layer \cite{krasnopolsky2012photochemical,marcq2013variations}. However, the dynamics driving such possibly periodic changes in the convective mixing {have} remained an open question till now \cite{marcq2018composition} and understanding variability on this interannual to decadal timescale is the focus of this paper. {The observational record for these long period variations is about 40 years long, spanning observations from Pioneer Venus} \cite{esposito1984sulfur}{ and Venus Express} \cite{marcq2013variations,marcq2020climatology}{ to the InfraRed Telescope Facility (NASA IRTF)} \cite{encrenaz2020hdo}.
 
Outgoing thermal flux from the hot, deep atmosphere is {largely} absorbed near the cloud-base and is the driver of convective activity in the middle cloud layer \cite{pollack1980greenhouse,lebonnois2015analysis}. Recent radiative transfer studies found that trace gas abundances near the cloud-base, in particular water abundance, have a large effect on the heating of the cloud-base \cite{lee2016sensitivity,haus2015radiative} which should affect the strength of convective mixing. In this paper, we develop a framework to investigate how changes in water abundance near the cloud-base affect vertical mixing within the
cloud, possibly giving rise to regular variations in water and sulfur-dioxide transport to cloud-tops on decadal timescales. Section \ref{sec:modeldesc} describes the model setup and Section 3 examines the solutions to the model and discusses their
implications as well as the limitations of the model. Section 4 summarizes the findings and suggests directions for future investigations. 

\section{Model Description}
\label{sec:modeldesc}

\begin{figure*}
\includegraphics[width=\columnwidth]{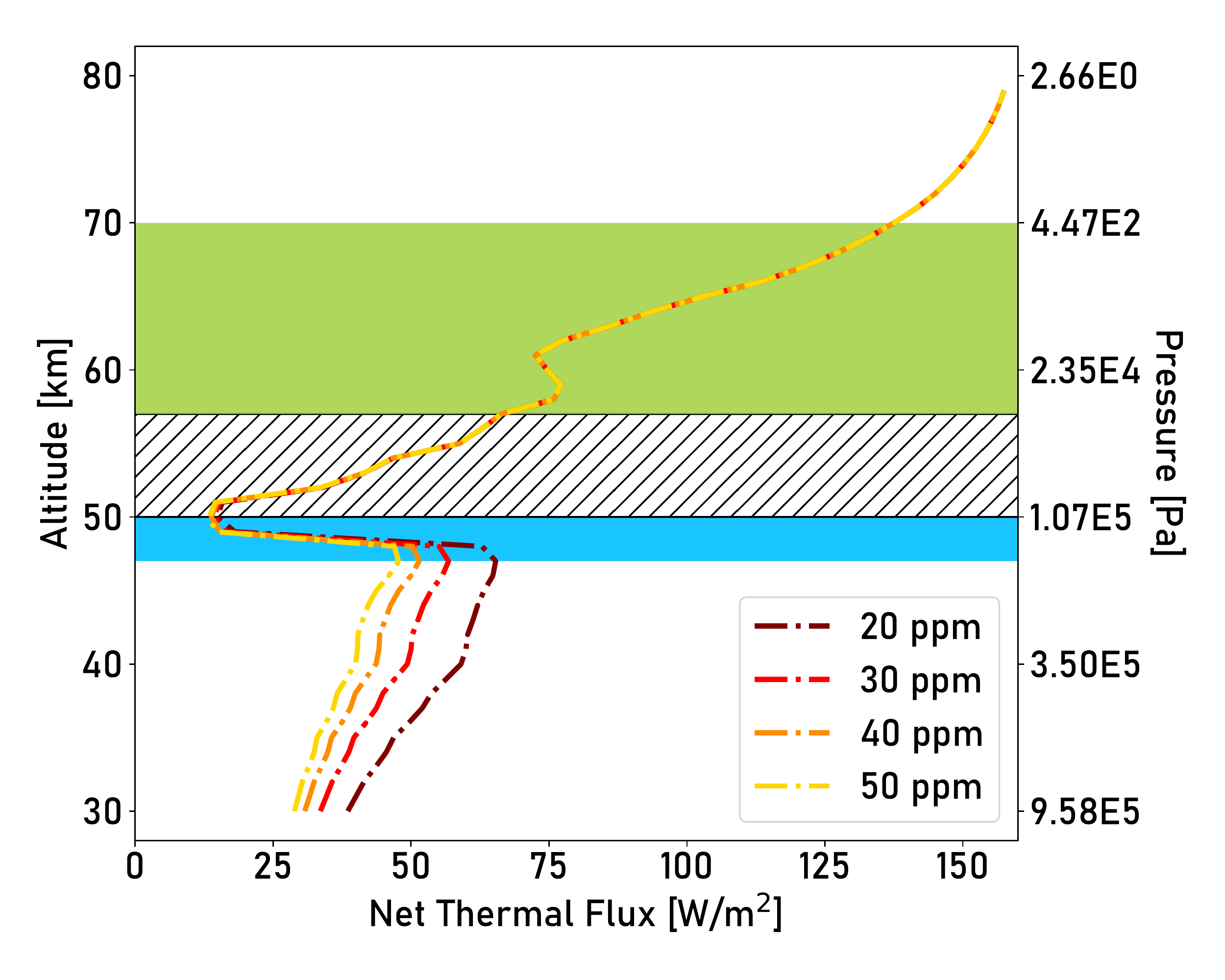}
  \caption{Simplified structure used in this work to represent the atmosphere of Venus. The green box shows the upper level dominated by photochemical clouds, the blue box shows the cloud-base and the hatched box in between shows the convective region. The thermal net flux profiles for different abundances of water in the blue box are from \citeA{lee2016sensitivity}, same as Fig 9c of that paper. }
  \label{fig:atmprofile}
\end{figure*}

 Our approach here will be to construct a simplified {box-}model to study variability within and below the cloud layer. In the following discussion, we describe the vertical regions of interest in the atmosphere as below \cite{titov2018clouds}:
\begin{enumerate}
\item Upper box is the photochemically dominated upper cloud region (57-70 km, green colored region in Fig. \ref{fig:atmprofile}). 
\item Middle box is the convectively unstable middle cloud region (50-57 km, hatched region). We will refer to this region as the convective cloud or convective column.
\item Lower box is the lower cloud where water abundance strongly influences thermal flux (47-50 km, blue colored region). We will refer to this region as the cloud-base.
\end{enumerate} 
  Note that by "convective cloud-top" in this paper, we refer to the boundary between the middle and upper boxes in Fig. \ref{fig:atmprofile}, while "cloud-top" will refer to the top of the upper box. The {abovelisted} vertical ranges are approximations, and the actual values of these altitudes change with time, {latitude and local time} as seen from various studies \cite[for e.g.]{tellmann2009structure,barstow2012models,imamura2014inverse}.
 \begin{table}
 \caption{List of Variables Used}
 \centering
 \begin{tabular}{l l}
 \hline
  Symbol  & Meaning  \\
 \hline
   $Q_R$  & IR heating at the cloud-base   \\
   $[H_2O]$  & Water abundance at the cloud-base   \\
   $[H_2O]^o$ & Water abundance at the cloud-base at equilibrium  \\
   $S_w$  & Sensitivity of IR heating to water abundance   \\
   $T$  & Temperature of cloud-base   \\
   $\Delta T$  & Change in temperature of cloud-base   \\
   $t$ & Time \\
   $\rho$  & Density of cloud-base   \\
   $c_P$  & Specific heat at constant pressure of the atmosphere at cloud-base   \\
   $L_b$  & Height of cloud-base   \\
   $L$ & Height of convective cloud layer \\
   $L^o$ & Height of convective cloud layer at equilibrium\\
   $\Delta L$ & Change in height of convective cloud layer \\
   $\Gamma _{ad}$ & Dry adiabatic lapse rate \\
   $[H_2SO_4]$ & Sulfuric acid abundance at the cloud-base \\
   $[H_2SO_4]^o$ & Sulfuric acid abundance at the cloud-base at equilibrium \\   
   $\Delta[H_2SO_4]$ & Change in sulfuric acid abundance at the cloud-base \\
   $\Delta[H_2O]$ & Change in water abundance at the cloud-base \\
   $K$ & Eddy diffusivity in the convective cloud and cloud-base \\
   $K^o$ & Eddy diffusivity in the convective cloud and cloud-base \\
   $H$ & Scale height of the atmosphere at the cloud base \\
   $[H_2O]^o_{deep}$ & Water abundance in the deep atmosphere (at 35 km) \\
   $t_{mix}$ & Chemical mixing timescale at cloud-base \\
   $v_{conv}$ & Vertical velocity in the convective cloud \\
   $\alpha$ & Constant of proportionality between $v_{conv}$ and $L$ \\
   $L_{mix}$ & Eddy mixing length \\
   $x$ & Non-dimensionalized water abundance anomaly at cloud-base \\
   $y$ & Non-dimensionalized convective layer height anomaly \\
   $a, b$ & Coefficients in the non-dimensionalized equations \\
   $T_{osc}$ & Time period of the oscillation \\
 \hline
 \multicolumn{2}{l}{Listed approximately in order of appearance in the text.}
 \end{tabular}
 \end{table}
 
\subsection{Variability of Cloud Level Convection}
\label{sec:upwardmix}
Our first step is to describe the dependence of cloud level convection to changes in cloud-base thermal flux. As demonstrated by the radiative transfer modeling of \citeA{lee2016sensitivity}, the water abundance in the lower box has a significant effect on the net thermal flux at the cloud-base (also see Fig.\ref{fig:atmprofile}). The vertical gradient of the thermal flux is the radiative heating at that level. Since extrema occur near the boundaries of the cloud-base, we approximate the radiative heating flux at the cloud-base as the difference between the maximum and minimum values of the thermal flux in the lower box. Using this definition, we calculate the heating fluxes from the profiles shown in Fig. \ref{fig:atmprofile} giving heating flux values of 47.29, 40.38, 36.08 and 33.07 $Wm^{-2}$ as water abundance is varied from 20, 30, 40 to 50 $ppm$. This is approximately linear, and a least squares linear fit for these values yields a sensitivity of the heating flux to the water abundance of -0.47  $Wm^{-2}ppm^{-1}$. We parameterize this dependence of cloud-base heating on water abundance as a linear relationship:

\begin{linenomath*}
\begin{equation}
Q_R([H_2O])-Q_R([H_2O]^o)= - S_w([H_2O]-[H_2O]^o)
\label{eqn:waterflux}
\end{equation}
\end{linenomath*}
where $Q_R([H_2O])$ is the cloud-base heating flux (in units of $Wm^{-2}$) as a function of lower box water abundance, denoted by $[H_2O]$ in units of parts per million ($ppm$), $Q_R([H_2O]^o)$ = 47 $Wm^{-2}$ is the cloud-base heating at an equilibrium water abundance of about $[H_2O]^o$ = 20 $ppm$ \cite{marcq2018composition} and $S_w$ = 0.47  $Wm^{-2}ppm^{-1}$ is the sensitivity of the heating flux to the water abundance.

We relate changes in this thermal heating flux to changes in convective strength. The thermal heating changes the lapse rate and thus the static stability of the atmosphere, which is then adjusted to an adiabatic lapse rate by convective mixing wherever the lapse rate is superadiabatic. Typically, such a radiative-convective equilibrium is calculated recursively by adjusting the height of the convective layer once the radiative fluxes are known till the atmosphere is either statically stable or neutral at all altitudes and solar and thermal fluxes are balanced \cite[Chap 4 $\&$ 5, for e.g.]{pierrehumbert2010principles}. Here, we take a simpler approach to estimate the strength of convection with the changes in radiative forcing. At the cloud-base, a change in thermal heating from the equilibrium value of $Q_R([H_2O]^o)$ to a new value of $Q_R([H_2O])$ will result in the cloud-base temperature relaxing radiatively according to \cite{spiegel1957smoothing}:
\begin{linenomath*}
\begin{align}
\frac{dT}{dt} &= \frac{Q_R([H_2O])-Q_R([H_2O]^o)}{\rho c_P L_b} 
\label{eqn:tempvar}
\end{align}
\end{linenomath*}
where $\rho \approx 2~kgm^{-3}$ is the atmospheric density at the cloud-base \cite{seiff1985models}, $c_P \approx 1000 JK^{-1}kg^{-1}$ is the specific heat \cite{lebonnois2010superrotation} and $L_b = 3~km$ is the height of the cloud-base. In the above equation, heat transport is assumed to be purely radiative, advective and eddy heat transport terms are neglected. This is an approximation, and can be justified on the basis that infrared radiation constitutes about 75-85$\%$ of the upward heat transport even within the convective cloud \cite{imamura2014inverse}, and this fraction should be even higher at the cloud-base where there is no convection and eddy diffusivity is smaller than in the convective layer \cite{woo1982small}. Thus{,} the above equation will have errors of order $10\%$, but it is sufficiently accurate to understand the nature of temperature variability in response to flux changes. 

We now explore the relationship between the convective strength in the clouds and the cloud-base temperature change. The cloud-base temperature changes only affect the temperature profile within the convective layer but are not expected to alter the thermal flux above the clouds due to the large infrared opacity of the lower and middle cloud layer \cite{lee2016sensitivity}. Since a significant fraction of thermal flux from the top of the convective layer escapes directly to space through the relatively more transparent upper clouds \cite{lebonnois2015analysis}, we make the assumption that the temperature at the top of the convective cloud remains a constant irrespective of cloud-base temperature so that the net thermal flux above the clouds is invariant. So, if the cloud-base temperature changes by an amount $\Delta T$, to keep the convective cloud-top temperature constant, the convective cloud-top height must change by an amount $\Delta L$ given by \cite{vallis2014lecture}:
\begin{linenomath*}
\begin{align}
\Delta L &=  \frac{\Delta  T}{\Gamma _{ad}}
\label{eqn:heightvar}
\end{align}
\end{linenomath*}
where $\Gamma_{ad} = 10^o~Kkm^{-1}$ is the adiabatic lapse rate in the convective clouds \cite{tellmann2009structure}. We will use the convective layer height as a measure of convective strength and its variability. Combining Eqns. \ref{eqn:tempvar} and \ref{eqn:heightvar}, we get a prognostic (i.e. an evolution) equation for the height of the convective cloud as a function of cloud-base heating:
\begin{linenomath*}
\begin{align}
\frac{dL}{dt} = \frac{(Q_R([H_2O])-Q_R([H_2O]^o))}{\rho c_P L_b \Gamma_{ad}}
\label{eqn:buoyprog}
\end{align}
\end{linenomath*}

\subsection{Vertical Transport of Water}
Chemical modeling of Venus has thus far focused primarily on calculating steady state abundances, since models which fully couple chemistry and dynamics still do not exist for Venus \cite{marcq2018composition,shao2020revisit}. Thus, in this section we start with results from such steady state chemical calculations and then couple the shifts between steady states to dynamical variability. Within the clouds, the fluxes of water and sulfuric acid are coupled together and their vapor pressures are in liquid-vapor equilibrium with a sulfuric acid-water mixture at steady state \cite{krasnopolsky1994h2o}. The downward eddy flux of sulfuric acid is equal to the production rate of sulfuric acid in the upper cloud and is given by the product of the eddy diffusivity and the vertical gradient of sulfuric acid abundance. Given a fixed rate of sulfuric acid production, the changes in the eddy diffusivity will alter the sulfuric acid gradient such that the downward eddy flux remains a constant \cite{krasnopolsky2015vertical}. Thus, cloud-base sulfuric acid abundances when chemical abundances have adjusted to changes in eddy diffusivity follow this relationship \cite{krasnopolsky2015vertical}:
\begin{linenomath*}
\begin{equation}
[H_2SO_4]K = [H_2SO_4]^oK^o 
\label{eqn:proportionality}
\end{equation}
\end{linenomath*}
where the right hand side values {are} at radiative equilibrium, and $K$ is the vertical eddy diffusivity at the cloud-base. We assume that the eddy diffusivity remains invariant with height within the convective clouds and the cloud-base \cite{krasnopolsky2012photochemical}. The implications of this assumption not being accurate are touched upon in Sec. \ref{subsec:solutions} and \ref{subsec:limitations}. Further, the assumption of a fixed sulfuric acid production rate implies that the upward fluxes of water and sulfur dioxide and the downward flux of sulfuric acid are also constant regardless of eddy diffusivity. This is a very strong assumption and it comes from using steady state chemical model results. Its caveats are further discussed in Sec. \ref{subsec:limitations}. Such a fixed flux condition usually implies that the gradients of the chemical abundance should follow a{n inverse} proportionality relationship with the eddy diffusivity ($Flux=K\frac{d[H_2SO_4]}{dt}=~constant$), but here the cloud-base abundances also follow this relationship, as expressed by Eqn. \ref{eqn:proportionality}. This is because the cloud-base abundances are the lower boundary values set by the gradient within the clouds, so for example, a doubling of the gradient will double the cloud-base abundance. 

The chemical abundance equilibrates to such changes in dynamics on a characteristic mixing timescale given by $t_{mix}=H^2/K^o$ \cite{krasnopolsky2012photochemical}, where $H = 5~km$ is the atmospheric scale height at the cloud-base. Then, the change in sulfuric acid concentration arising from a change in eddy diffusivity can be written using Eqn. \ref{eqn:proportionality} as:
 \begin{linenomath*}
\begin{equation}
\Delta [H_2SO_4] = [H_2SO_4] - [H_2SO_4]^o = \frac{[H_2SO_4]}{K^o}\left(K^o-K\right) 
\label{eqn:hsovar}
\end{equation}
\end{linenomath*}
   In the Krasnopolsky cloud models \cite{krasnopolsky1994h2o,krasnopolsky2015vertical} which calculate the abundances and fluxes of water and sulfuric acid within the clouds, the sum of water and sulfuric acid mixing ratios at the cloud-base (and below) are constrained to sum to a constant (equal to the deep atmospheric water mixing ratio) due a hydrogen element conservation constraint. The hydrogen element conservation is the condition that since water and sulfuric acid are the dominant hydrogen bearing species, and there is no significant loss of hydrogen (for example, to atmospheric escape or condensation), changes in the abundance of water and sulfuric acid must balance the upward and downward fluxes of hydrogen such that the total abundance of hydrogen at each vertical level remains a constant. Therefore, the change in the cloud-base water concentration has the same magnitude but opposite sign as that of the cloud-base sulfuric acid. These constraints give us two more relationships:
\begin{linenomath*}
\begin{align}
\Delta [H_2SO_4] = -\Delta [H_2O] \\
[H_2SO_4] + [H_2O] = [H_2O]^{o}_{deep}
\end{align}
\end{linenomath*}
where $[H_2O]^{o}_{deep}$ is the deep atmospheric water abundance where no sulfuric acid exists in the atmosphere, around 35 km altitude.
Then, we can write Eqn. \ref{eqn:hsovar} in terms of water abundances using the above two relations as:
\begin{linenomath*}
\begin{align}
\Delta [H_2O] &= -\left(\frac{[H_2O]^{o}_{deep}-[H_2O]}{K^o}\left(K^o-K\right) \right) \\
&= \frac{[H_2O]^{o}_{deep}-[H_2O]}{K^o}\left(K-K^o\right)
\end{align}
\end{linenomath*}
  Since the chemical compositions relax in response to dynamical perturbations over the mixing timescale $t_{mix}$, we can construct a simple prognostic equation for the evolution of water abundance:
\begin{linenomath*}
\begin{align}
\frac{d[H_2O]}{dt} &= \frac{\Delta[H_2O]}{t_{mix}} \\
&= \frac{[H_2O]^{o}_{deep}-[H_2O]}{K^o}\left(K-K^o\right)*\left(H^2/K^o \right)^{-1} \\
&= \frac{[H_2O]^{o}_{deep}-[H_2O]}{H^2}\left(K-K^o\right)
\label{eqn:waterprog}
\end{align}
\end{linenomath*}

\subsection{Closure Condition and Coupled Model Equations}
\label{subsec:closure}
We have found two prognostic equations and three unknowns - water abundance, convective layer height and the eddy diffusivity. Thus, we need a closure condition linking the convective layer height and the eddy diffusivity to solve this system of equations. Turbulence-resolving numerical simulations of Venus' atmosphere found that the convective vertical velocity doubles for a doubling of the convective layer height \cite{lefevre2018three}. This corresponds to a constant convective timescale, making the buoyant acceleration increase proportionally to height of the convective layer, and thus we can write a linear scaling relationship between convective velocity and convective layer height:
\begin{linenomath*}
\begin{equation}
v_{conv} = \alpha L
\end{equation}
\end{linenomath*}
Convective velocities are of order $1~ms^{-1}$ \cite{blamont1986implications} and convective layer height at equilibrium is $L^o = 7~km$ (as defined in Sec \ref{sec:modeldesc}), so the constant of proportionality $\alpha=1.4\times10^{-4}~s^{-1}$. This value is the inverse of the convective timescale. The eddy diffusivity can then be written as
\begin{linenomath*}
\begin{equation}
K = v_{conv}*L_{mix} = \alpha LL_{mix}
\end{equation}
\end{linenomath*}
 where $L_{mix}$ is a mixing length. Note that we have not made a distinction between thermal and momentum eddy diffusivities and the above value is derived from thermal considerations (layer height depends on cloud-base temperature, as defined in Sec. \ref{sec:upwardmix}) but is used for estimating dynamical mixing of chemical species. Since the Prandtl number, which is the ratio of the momentum to the thermal diffusivity is of order unity and does not vary much with altitude from 45-60 $km$ on Venus \cite{morellina2020global}, using a single value of eddy diffusivity for both is a crude but acceptable simplification. \add[referee2]{The relationship derived above indicates that higher thermal heating flux at the cloud base (represented by a larger convective layer height) leads to a higher eddy diffusivity, which is consistent with other numerical studies of eddy mixing in the Venus cloud layer} \cite{yamamoto2014idealized}.
 
 Estimating the appropriate magnitude for both K and L$_{mix}$ is not straightforward. As summarized in the recent work of \citeA{bierson2019chemical}, dynamical studies such as cloud microphysics modeling \cite{imamura2001microphysics,mcgouldrick2007investigation} support high values of K in the range of $10^2-10^3~m^2s^{-1}$, whereas chemical models \cite{krasnopolsky2012photochemical,krasnopolsky2015vertical} prefer values of $1~m^2s^{-1}$ or lower to prevent large excess transport of trace gases such as sulfur dioxide to the upper atmosphere. \citeA{bierson2019chemical} argue that this inconsistency indicates a gap in our understanding of the factors affecting chemical transport in the clouds, such as an unknown chemical sink or cloud interactions. Since our approach to modeling requires an eddy diffusivity consistent with chemistry and chemical mixing timescales, we employ $K_{zz} = 1~m^2s^{-1}$ \cite{krasnopolsky2012photochemical}, which gives an $L_{mix}=1~m$. We acknowledge that this $L_{mix}$ is much smaller than the $\sim1~km$ values calculated with the VEGA balloon observations \cite{blamont1986implications}.
 
We can write the change in eddy diffusivity in terms of the change in layer height as:
\begin{linenomath*}
\begin{align}
\Delta K = K - K^o = \alpha L_{mix}(L - L^o)
\label{eqn:vertbuoy}
\end{align}
\end{linenomath*}

Substituting Eqn \ref{eqn:waterflux} into Eqn \ref{eqn:buoyprog} and Eqn \ref{eqn:vertbuoy} into Eqn \ref{eqn:waterprog}, we end up with a coupled set of equations for water abundance and convective layer height:

\begin{linenomath*}
\begin{align}
\frac{d[H_2O]}{dt} &= \frac{\alpha L_{mix}(L-L^0)([H_2O]^o_{deep}-[H_2O])}{H^2} \label{eqn:finalwater} \\
\frac{dL}{dt} &= -\frac{S_w([H_2O]-[H_2O]^o)}{\rho c_P L_b \Gamma_{ad}} \label{eqn:finalheight}
\end{align}
\end{linenomath*} 
Given a pair of perturbed initial conditions for $[H_2O]$ and $L$, these equations can be integrated to give solutions for water abundance and convective strength as functions of time. These equations describe the effects of tendencies of both temperature and cloud-base water to relax to equilibrium values. Since these tendencies are coupled, the resulting behavior is an oscillation instead of first-order approach to equilibrium. We explore the nature of the solutions in the following section.
\section{Results and Discussion}
\subsection{Solutions to the Coupled Equations}
\label{subsec:solutions}
 The model equations are structurally very similar to the set obtained by \citeA{yano2012finite} [henceforth YP12] in their studies of convective cycles and represent a recharge-discharge oscillator. We note that while the equations have a similar structure,
the present derivation represents very different physics on different
timescales. However, in a mathematical sense they both involve similar coupled interactions, for example, the cloud base mass flux in YP12 and $[H_2O]$ in the present model are both being driven by departures in the cloud work
function in YP12 and the depth of the convective layer here respectively. Furthermore, where in YP12 the cloud work function declines with {an} increase in the cloud base mass flux as buoyancy is released, the present model has the convective layer height decreasing with an increase in water abundance owing to reductions in the heating rate.  As a result, we obtain similar dynamics, albeit on quite
different timescales. We make the following substitution to non-dimensionalize the present equations: 
\begin{linenomath*}
\begin{align}
[H_2O] &= [H_2O]^o(1+x) \label{eqn:waternon} \\
L &= L^o(1+y) \label{eqn:heightnon} 
\end{align}
\end{linenomath*} 
We take $[H_2O]^{o}_{deep}$ = $1.5*[H_2O]^o$ = 30 $ppm$   \cite{marcq2018composition} and $L^o$ = 7 $km$, as before. Variables $x$ and $y$ will be referred to as the non-dimensional cloud-base water abundance anomaly and the convective layer height anomaly respectively. Thus, the Eqns. \ref{eqn:finalwater} and \ref{eqn:finalheight} become:
\begin{linenomath*}
\begin{align}
\frac{dx}{dt} &= a(\frac{1}{2}-x)y \label{eqn:couplednon1} \\
\frac{dy}{dt} &= -bx \label{eqn:couplednon2}
\end{align}
\end{linenomath*} 
We can estimate the magnitudes of the coefficients $a$ and $b$ as follows:
\begin{linenomath*}
\begin{align}
&a = \frac{\alpha L_{mix} L^o}{H^2} = \frac{10^{-4}*1*10^4}{10^7} \approx 10^{-7}s^{-1} \approx (120~days)^{-1} \\
&b = \frac{S_w[H_2O]^o}{\rho c_P L_b L^o \Gamma_{ad}} = \frac{1*10}{1*10^3*10^3*10^4*10^{-2}} \approx 10^{-7}s^{-1} 
\end{align}
\end{linenomath*}
The coefficients have clear physical meanings: $a$ is the inverse of the characteristic eddy mixing timescale in the convective clouds and $b$ is the inverse of the radiative cooling timescale of the convective column, since it takes the form of the thermal flux over the heat capacity of the column. {If the value of $K$ is very different from the set value of $1~m^2s^{-1}$, for example $0.1~m^2s^{-1}$ and $100~m^2s^{-1}$, the constant $a$ will become approximately $(3~years)^{-1}$ and $(1~day)^{-1}$ respectively.}

The steady-state of the model in Eqns. \ref{eqn:couplednon1} and \ref{eqn:couplednon2} is $x=y=0$. The characteristic oscillation timescale for small perturbations to the steady of this coupled set of equations can be estimated from the linearized form of the Equations \ref{eqn:couplednon1} and \ref{eqn:couplednon2}:
\begin{linenomath*}
\begin{equation}
\frac{dx}{dt} = \frac{a}{2}y , \frac{dy}{dt} = -bx
\label{eqn:linearized}
\end{equation}
\end{linenomath*}
The above two equations can be combined to yield the equation for a simple harmonic oscillator in either variable
\begin{linenomath*}
\begin{equation}
\frac{d^2}{dt^2} \left( x,y\right) = - \frac{ab}{2} (x,y)
\label{eqn:simpharm}
\end{equation}
\end{linenomath*}
The time period of oscillation is given by 
\begin{linenomath*}
\begin{equation}
T_{osc} =  2\pi\sqrt{\frac{2}{ab}} \approx 3-9~years
\label{eqn:timeperiod}
\end{equation}
\end{linenomath*}
with the 3 year period for $[a] = 10^{-7}s^{-1}$ {($K = 1~m^2s^{-1}$)} and 9 year period for $[a] = 10^{-8}s^{-1}$ {($K = 0.1~m^2s^{-1}$)}, $[b] = 10^{-7}s^{-1}$ is not changed. We vary $a$ since our simplifying assumption of a height invariant eddy diffusivity is not always true. The eddy diffusivity near the cloud-base (45 $km$) was estimated from observations to be an order of magnitude smaller than near the top of the convective clouds (60 $km$) \cite{woo1982small}. Shorter mixing lengths, corresponding to smaller values of the eddy diffusivity, signify inefficient mixing and allow for the development and persistence of
larger vertical concentration gradients and slower oscillations as a result. 

Since the above is a rather large range for the oscillation period, a natural question to ask is: what is the appropriate timescale to consider when using this model to interpret observations of Venus?  We note that the oscillation period is the geometric mean of the eddy mixing and radiative cooling timescales of the convective column multiplied by a constant {value of 8.8}. The {radiative} cooling timescale is fairly well understood since remote sensing and associated radiative modeling of Venus has a long heritage, but the mixing timescale near the cloud-base could very well vary by an order of magnitude in either direction and is not strongly constrained by continuous observations, since the last observationally constrained values of eddy diffusivity are from the 1980s (for e.g., see Fig. 3 of \citeA{bierson2019chemical}). At this time, we cannot make a stronger prediction than to say the timescale is roughly interannual to decadal. \citeA{encrenaz2020hdo} {estimate that sulfur dioxide abundances at the cloud top decrease from a peak to a minimum value in about 5-7 years from 2008 to 2015, which is roughly consistent with the model timescale.}

Furthermore, since Eqn. \ref{eqn:simpharm} does not have any damping or growing terms, oscillations resulting from an initial perturbation will maintain both their amplitude and period indefinitely. This is also true for the nonlinear set of equations (Eqns. \ref{eqn:couplednon1} and \ref{eqn:couplednon2}). As we will see, its solutions persist indefinitely, giving rise
to sustained oscillations around the steady state. The linearized
system always has pure imaginary eigenvalues, since both $a$ and
$b$ must be positive. Therefore the qualitative behavior of the nonlinear system cannot be deduced from its linear approximation, because the conditions for equating their qualitative behaviors are not met. Eigenvalues of the linearized system are pure imaginary, and therefore not hyperbolic (non-zero real part) as required by the result of Hartman and Grobman giving conditions for the nonlinear phase portrait to qualitatively follow from the linearized one \cite{guckenheimer2013nonlinear}.

Still it is possible to infer the behavior of the nonlinear system
from explicit integration since, from Eqns. \ref{eqn:couplednon1} and \ref{eqn:couplednon2}:
\begin{linenomath}
\begin{equation}
aydy=b\left\{ 1-\frac{\frac{1}{2}}{\frac{1}{2}-x}\right\} dx
\end{equation}
\end{linenomath}
which is integrated for
\begin{linenomath}
\begin{equation}
ay^{2}-b\left\{ 2x+\log\left(\frac{1}{2}-x\right)\right\} =C
\label{eqn:Cexp}
\end{equation}
\end{linenomath}
where $C\left(x_{0},y_{0}\right)=ay_{0}^{2}-b\left\{ 2x_{0}+\log\left(\frac{1}{2}-x_{0}\right)\right\} $
is constant in time, depending only on initial conditions $\left(x_{0},y_{0}\right)$. Such "integrability" was demonstrated earlier for the recharge-discharge
oscillator model of YP12. 

\begin{figure*}
  \includegraphics[width=\columnwidth]{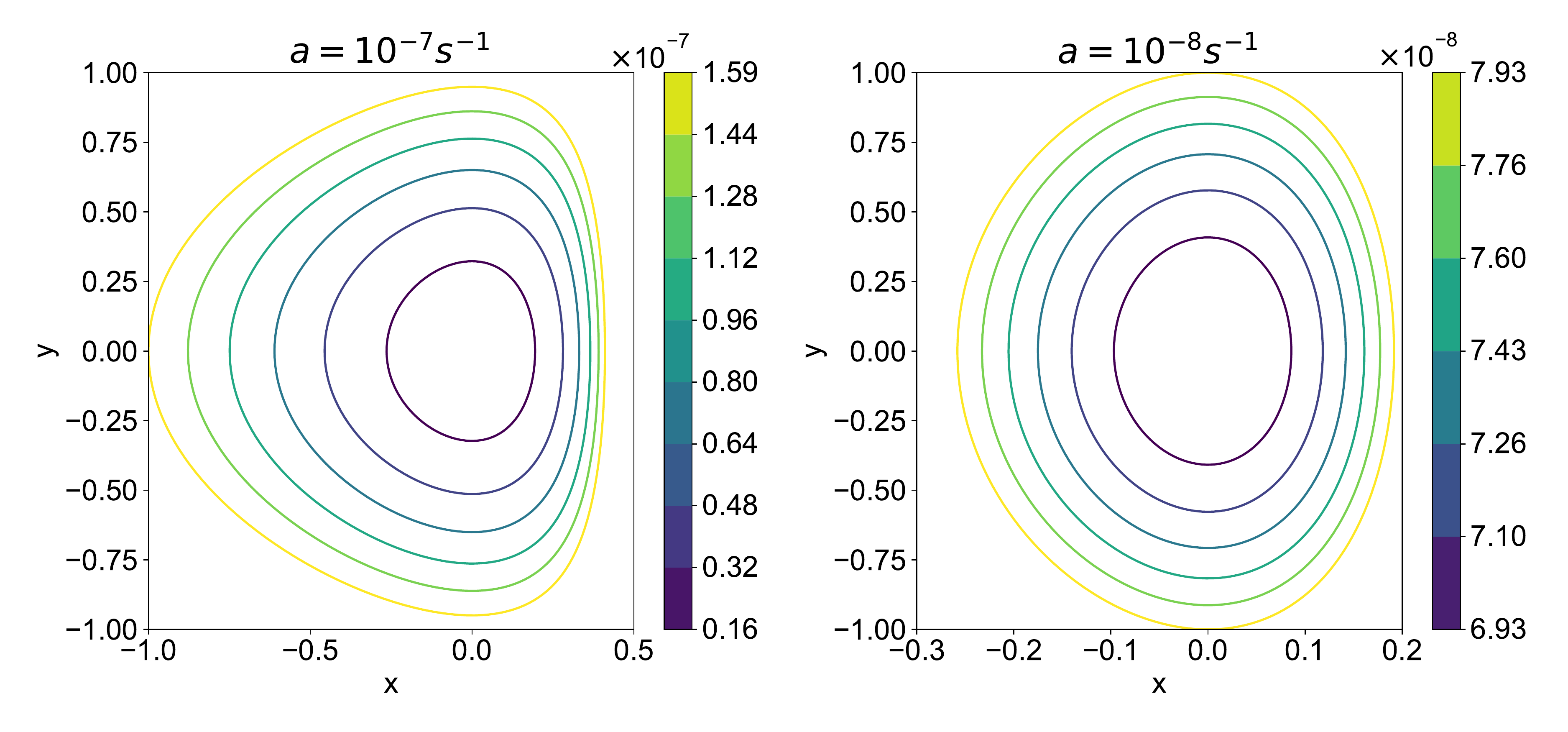}
  \caption{Phase plots showing the closed contours of the conserved quantity $C$ for periodic solutions of the nonlinear model equations, where $x$ and $y$ are the non-dimensional cloud-base water abundance anomaly and convective layer height anomaly respectively. The parameters are set to $[b]=10^{-7}s^{-1}$, while $a$ is varied as shown in the panel title.}
  \label{fig:phaseplot}
\end{figure*}

The present model's dynamics depends on function $f\left(x\right)=-b\left\{ 2x+\log\left(\frac{1}{2}-x\right)\right\} $
appearing in the above expression, which takes minimum value at $x=0$
where its derivative $f'\left(x\right)$ vanishes. Everywhere, this
function has positive second derivative $f''\left(x\right)=b\left(\frac{1}{2}-x\right)^{-2}>0$,
so it is convex. As a result $C\left(x,y\right)$ is convex in both $x$
and $y$, its contours are given by closed curves, and solutions of
the nonlinear equations become periodic as show in Fig. \ref{fig:phaseplot}, with
their time-evolution always conserving $C\left(x_{0},y_{0}\right)$.
Large-amplitude oscillations are quite asymmetric in the left panel,  because unlike the symmetric quadratic potential for $y$, the logarithmic term makes $f\left(x\right)$ rise sharply as $x$ approaches $1/2$. We also note that the minimum values of $x$ and $y$ have a lower bound at $-1$. This is because the cloud-base water abundance and the lower height cannot be smaller than zero physically (Eqns. \ref{eqn:couplednon1} and \ref{eqn:couplednon2}). Thus, the right panel remains close to the small-amplitude limit for $x$, since these values are already sufficient to cause large oscillations in $y$, of unit amplitude.

\begin{figure*}
  \includegraphics[width=6cm]{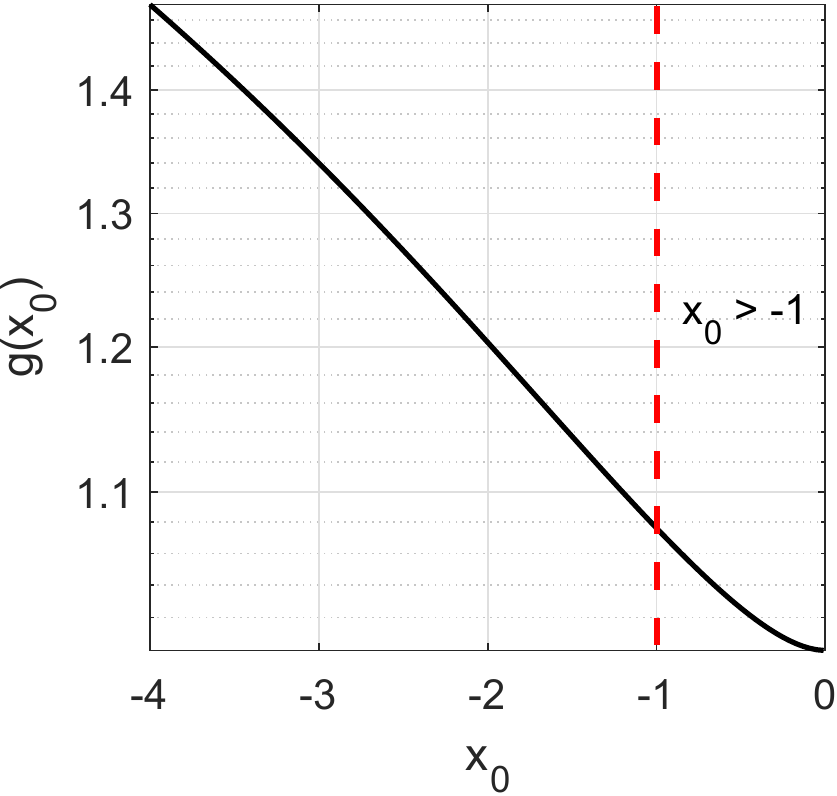}
  \caption{Function $g$ showing the time period increase with respect to the simple harmonic period as a function of the minimum cloud-base water abundance anomaly, $x_0$. Physically speaking, the model is only valid to the right-side of the red line, such that the total cloud-base water abundance ($[H_2O]^o(1+x)$) always remains positive.}
  \label{fig:periodamp}
\end{figure*}

The period of oscillation of the nonlinear system is found by integrating Eqn. \ref{eqn:couplednon1}:
\begin{linenomath}
\begin{equation}
T_{osc}=\int_{0}^{T}dt=2\int_{x_{0}}^{x_{1}}\frac{dx}{a\left(\frac{1}{2}-x\right)y}
\end{equation}
\end{linenomath}

where $x_{0}$ and $x_{1}$ are points of intersection between the
particular trajectory and the $x$-axis, for which $y=0$. Substituting
for $y$ from Eq. \ref{eqn:Cexp} we obtain the period 
\begin{linenomath}
\begin{equation}
T_{osc}\left(x_{0}\right)=2\pi\sqrt{\frac{2}{ab}}g\left(x_{0}\right)
\end{equation}
\end{linenomath}

where the first part is identical to that of the linearized model
in Eqn. \ref{eqn:timeperiod} and the second factor absorbs the integral
\begin{linenomath}
\begin{equation}
g\left(x_{0}\right)=\int_{x_{0}}^{x_{1}\left(x_{0}\right)}\frac{dx}{\sqrt{2}\pi\left(\frac{1}{2}-x\right)\sqrt{2\left(x-x_{0}\right)+\log\frac{\frac{1}{2}-x}{\frac{1}{2}-x_{0}}}}
\end{equation}
\end{linenomath}
Fig. \ref{fig:periodamp} shows the factor $g\left(x_{0}\right)$
that increases the period for oscillations having large amplitude. In summary, the period of oscillation of the nonlinear system varies
directly with the geometric mean of the eddy mixing timescale and
the radiative cooling timescale, and grows weakly with the amplitude
of the oscillation as measured by $x_{0}$. {The model oscillations conserve the quantity C(x,y) and small volume elements are recovered during the course of one period of the oscillation. The growth of volume elements is governed by the divergence of the vector field in} Eqns. \ref{eqn:couplednon1} and \ref{eqn:couplednon2} {and this, being equal to -ay, has a mean value of zero over a period of the oscillation. Therefore, this simple model cannot give rise to more complex dynamical behaviour such as chaos, which requires a source of dissipation to compensate nonlinear growth.}

\begin{figure*}
  \includegraphics[width=\columnwidth]{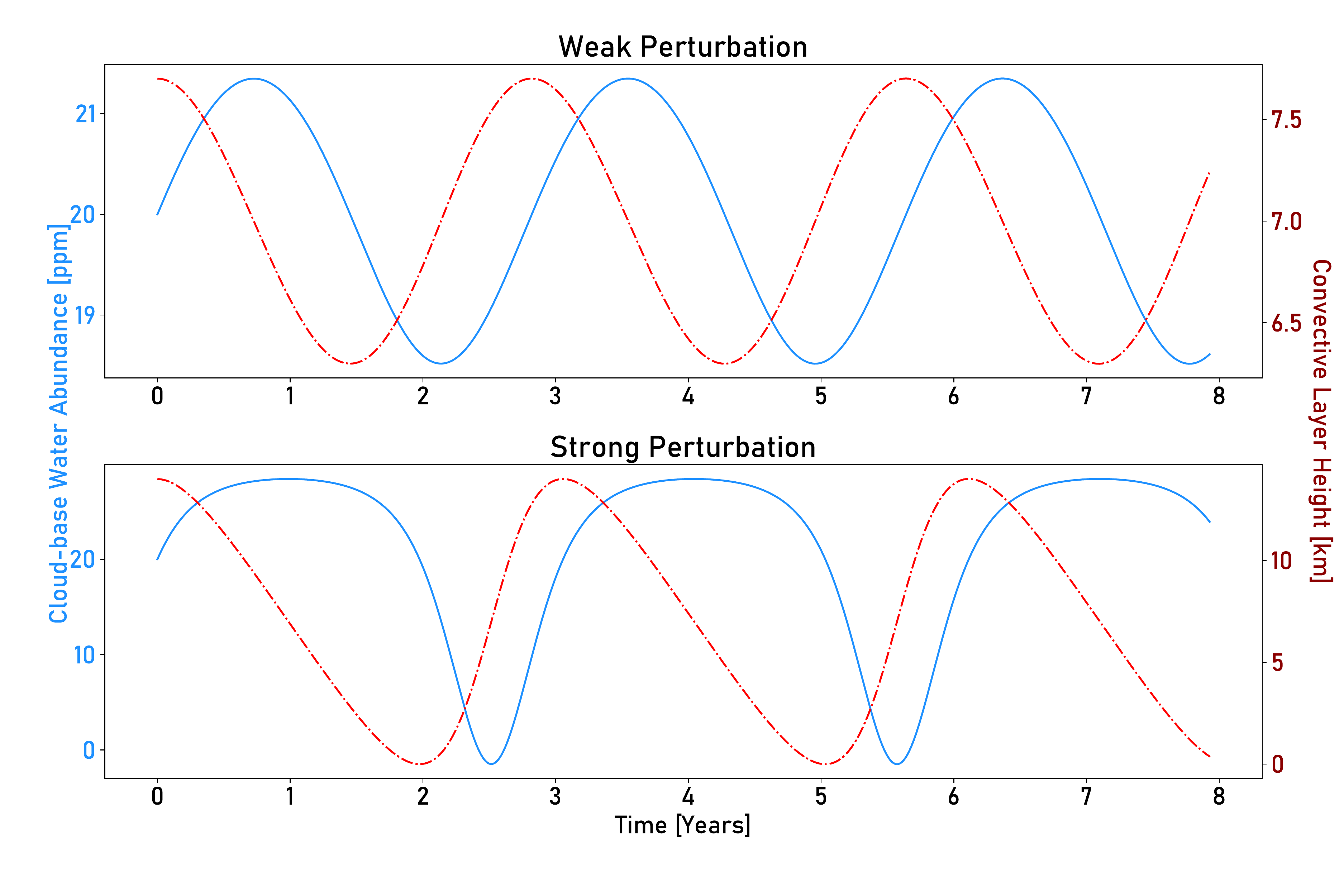}
  \caption{A simple first-order Eulerian scheme integration of the Equations \ref{eqn:couplednon1} and \ref{eqn:couplednon2} with $[a,b] = 10^{-7}s^{-1}$ showing interannual oscillations. The integrations were initialized with $[x,y]=[0,0.1]$ for the top panel and $[x,y]=[0,1]$ for the bottom panel. The blue solid line shows the cloud-base water abundance ($[H_2O]$), while the red dashed line shows the convective layer height ($L$). In the weak perturbation limit, the oscillations are harmonic and in the strong perturbation limit, they can be described as recharge-discharge oscillations.}
  \label{fig:oscillationsann}
\end{figure*}

\begin{figure*}
  \includegraphics[width=\columnwidth]{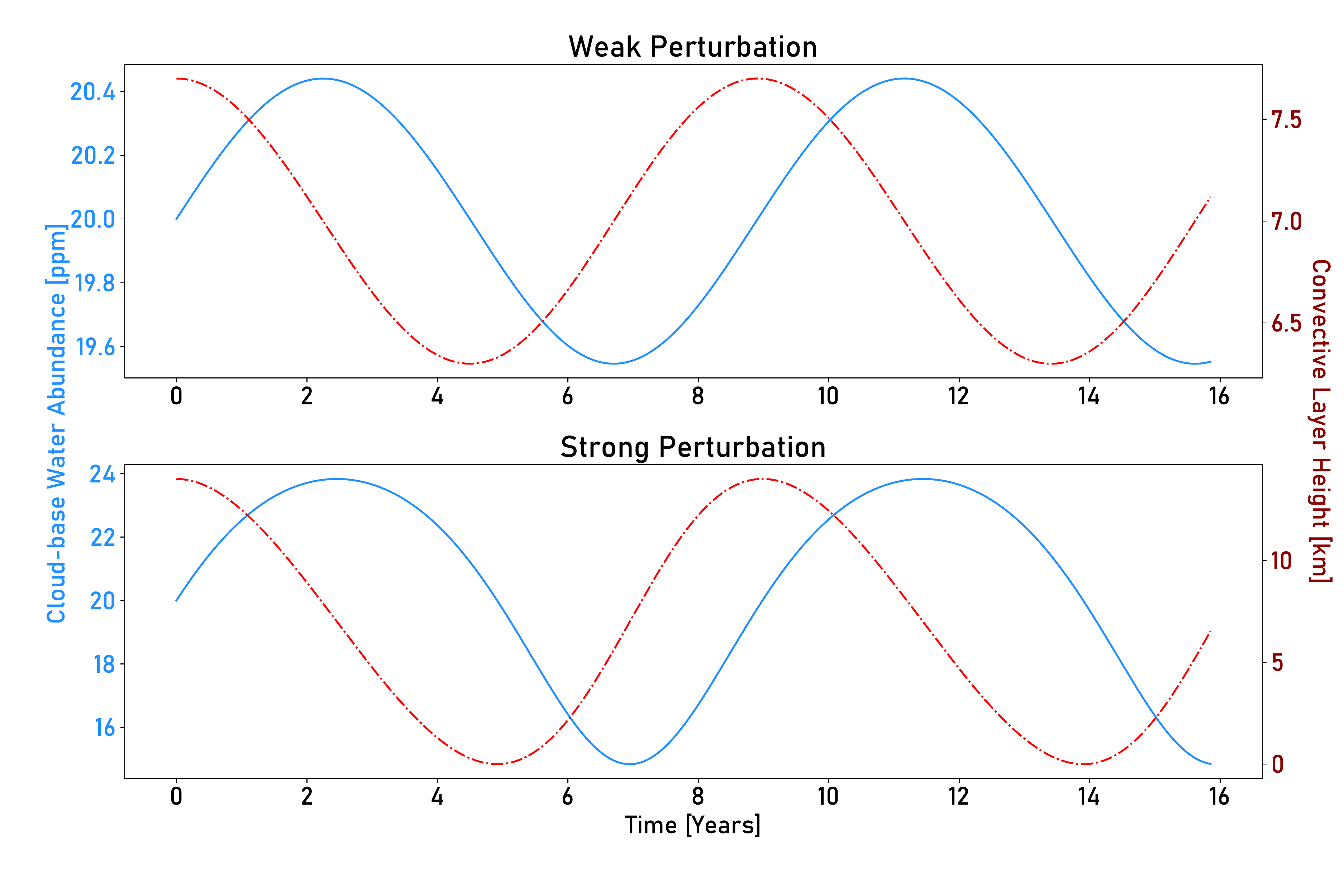}
  \caption{Same as Fig. \ref{fig:oscillationsann} but with $a = 10^{-8}s^{-1}$, showing decadal oscillations. As seen before in Fig. \ref{fig:phaseplot}, large changes in convective layer height produce only small changes in water abundance when the chemical mixing timescale is long. Thus, both the strong and weak perturbation limits behave similarly, like a harmonic oscillator.}
  \label{fig:oscillationsdec}
\end{figure*}

Integrations of the model equations are shown in Figs. \ref{fig:oscillationsann} and \ref{fig:oscillationsdec} for small and large perturbations. {The magnitude of changes in the linear regime are quite small and maybe difficult to observe. But the linear limit is nonetheless useful to derive the timescales of the oscillation and develop a physical intuition for how the coupled system of equations behaves in its simplest form.} In the non-linear limit the oscillations can be described as a recharge-discharge system: during the recharge phase the water abundance is nearly constant as the convective layer height decreases slowly. The discharge phase begins as the convective layer height approaches its minimum, during which time the cloud base water abundance begins to decrease rapidly. Water abundance reaches its minimum value and grows rapidly to reach its maximum value shortly after the convective layer height has reached its maximum.

\subsection{Relationship to observed sulfur dioxide oscillations}
Sulfur dioxide is transported to the cloud-tops from the cloud-base by convective transport in the middle clouds upto about 57 km (though this height is variable as discussed earlier) and diffusive transport in the upper clouds upto 70 km.  
The sulfur dioxide abundances at the cloud-tops having a strong relationship to the convective mixing strength (as well as sulfur dioxide and water abundances at the cloud-base) {has} been established in multiple chemical modeling studies \cite{krasnopolsky2012photochemical,krasnopolsky2018disulfur,
parkinson2015photochemical,bierson2019chemical}. The dependence of cloud-top sulfur dioxide on water abundance is complex, with correlated and anti-correlated behavior manifesting depending on whether water or sulfur dioxide is the relatively more abundant species at the convective cloud-top \cite{shao2020revisit}.

Our model indicates that water abundance and convective strength vary out of phase by about a quarter cycle of the oscillation in the weak perturbation limit, it becomes necessary to consider fine vertical resolution to accurately model the upward transport under such changing conditions. Furthermore, we have used a very strong simplifying assumption in Sec 2.2 that the sulfuric acid production rate remains a constant in the upper atmosphere, which implies that water and sulfur dioxide fluxes to the cloud-top are a constant. Thus, variations in water and sulfur dioxide abundances at all altitudes in the cloud only occur in response to changes in eddy diffusivity to keep the net flux a constant. In reality, there is no reason that these fluxes should be a constant as convective strength varies, thus our model cannot be directly used to quantitatively predict changes in these fluxes. 

{In spite of these limitations, we can roughly discuss the changes in sulfur dioxide expected as a result of the oscillations described above.} \citeA{krasnopolsky2012photochemical} {showed that as the convective layer top is moved by 10 $km$ (from 55 to 65 $km$), the sulfur dioxide abundance at the cloud top varied by a factor of 30. For the strong perturbations shown in} Figs. \ref{fig:oscillationsann} and \ref{fig:oscillationsdec}, {the change in layer height is about 14 $km$, which is slightly larger than that range. Thus, we would expect sulfur dioxide to vary between one and two orders of magnitude as a result of the convective layer height oscillations. This is in reasonable agreement with the observed magnitude of variability} \cite{marcq2013variations,marcq2020climatology,encrenaz2020hdo}. Stronger quantitative arguments would require a more thorough modeling of atmospheric dynamics and chemistry and is a promising direction for future work.

{Until now, the interannual variability of sulfur dioxide at the cloud tops was proposed to have two possible explanations: changes in vertical mixing due to atmospheric oscillations or volcanic injections of trace gases into the atmosphere} \cite{encrenaz2020hdo,marcq2020climatology}. {While the former explanation was preferred by some researchers due to Occam's razor} \cite{marcq2013variations,krasnopolsky2012photochemical} {, no mechanism was put forward to explain the existence of such an atmospheric oscillation. The model we described above provides just such a dynamical mechanism for interannual oscillations in the cloud layer mixing.}

\subsection{Model Predictions and Observables}

{The model predicts that the cloud-base water abundance and convective layer height should vary on interannual  timescales. Furthermore, the model also predicts that the extrema in water abundance lag behind the extrema in convective layer height} (see Figs. \ref{fig:oscillationsann} and \ref{fig:oscillationsdec}). {In the linear limit, the extrema in the two variables are separated by approximately 0.5 and 2 years for oscillations of period 3 and 9 years respectively. If this phase relationship does not hold, for example, if the water abundance peaks before the convective layer height peak, then the model introduced in this work can be falsified. In this section, we describe the nature of observations required to test these predictions.}

{The convective layer height can be observed using radio measurements, where the convective layer shows a distinctive adiabatic lapse rate as seen in observations from both Venus Express and Akatsuki} \cite[for e.g.,]{tellmann2009structure,imamura2017initial}. {A multi-year observational record of convective layer height would provide direct evidence of the oscillations proposed in this work. However, a direct analysis to extract possible interannual variations of the convective layer might be challenging, considering their limited spatial and temporal sampling coverages, which may be insufficient to distinguish latitudinal and local time dependences. Observations of water vapor abundance near the cloud-base would be the other way to check the predictions of the model described here. The 2.3 $\mu m$ band is used to retrieve water vapor abundance in the 30-45 $km$ sub-cloud altitude range on Venus. This region has been observed by several researchers with ground-based telescopes and space-based instruments such as VIRTIS on Venus Express} \cite{marcq2008latitudinal,tsang2008tropospheric,arney2014spatially}. {However, these studies have mostly focused on spatial variability and not on temporal variability. As shown in} 
Figs. \ref{fig:oscillationsann} and \ref{fig:oscillationsdec}, {the cloud-base water abundance variability can have magnitudes of 10-20 $ppm$ over a few years, which is much larger than observational uncertainties of order $\pm$2-4 $ppm$} (see Table 1 of \citeA{marcq2018composition} for a summary). 

{But the tricky issue about water vapor abundance retrieval at the 2.3 $\mu m$ window is that there are no long-term baseline observations up to now. Even VIRTIS on Venus Express yielded only three years of data. Additionally, the weighting functions in the 2.3 $\mu m$ band (which show the contribution function of thermal emission along altitudes to the observed signal) were rather broad, allowing limited vertical resolution of about 10 $km$} \cite{haus2015lower}. {Furthermore, the 2.3 $\mu m$ band contains spectral contributions from clouds, carbon monoxide, water, OCS and sulfur dioxide} \cite{marcq2008latitudinal}. {In particular, all of the water abundance retrieval studies are affected by uncertainties in cloud aerosol properties, which are handled differently in each study} \cite{marcq2008latitudinal,tsang2010correlations,arney2014spatially}. {Thus, isolation of temporal trends across multiple studies is hard to achieve at this moment.}
 
 {A long-term baseline analysis will be possible in nearthe  futurewith the proposed ESA mission Envision. The radio science measurements would obtain convective layer height information, and the VenSpec suite of spectrographic instruments intends to observe water and sulfur dioxide above and below the clouds}  \cite{ghail2017envision}. {In addition to the data from the previous missions, these high resolution measurements will have a sufficiently long time baseline capable of determining the existence and dominant periods of such oscillations. The ISRO Venus mission Shukrayaan-1} \cite{haider2018indian,isro2018shukrayaan}, {planned to launch in 2023, has spectrometers for studying atmospheric composition and a radio science instrument. Relevant data on water abundance and convective layer height may also come from this mission, but final instrument specifications are not publicly available at this time.}

\subsection{Model Limitations}
\label{subsec:limitations}
In our model, we have idealized or simplified many complex processes. In this section, we explore the implications of our simplifying assumptions breaking down. Firstly, the model is based upon perturbations from a stable equilibrium existing at a water abundance of $[H_2O]^o$ and convective layer height $L^o$. If, for example, the  cloud-base water abundance is fixed to some non-equilibrium value by processes not included within our model, Eqn. 2 shows that the cloud-base temperature will either {continuously} rise or fall, tending to unphysical values over time. In reality, a new temperature equilibrium will be reached, but our simple model cannot search for such a new equilibrium as a more complex numerical model could. {Thus, what we have treated as a closed system in our simplfied model is in reality an open system, and outside influences can force or damp these natural oscillations. Forcings can be caused by perturbations to water abundance by supply of volatiles from continuous or sporadic volcanic outgassing} \cite{esposito1984sulfur,smrekar2012constraints}{or changes in solar heating due to secular albedo changes over interannual timescales affecting convective layer height} \cite{lee2019long}. {Damping can be caused by heating and cooling influences on convection by changes in large scale circulation} \cite{lefevre2018three}{ or diffusive smoothing of water abundance anomalies by meridional circulation or vertical mixing in the deep atmosphere about which there is not much observational data} \cite{sanchez2017atmospheric}. {The typical timescales for these processes are a few years, though the rate of volcanic outgassing is not well constrained and its effects on the atmosphere could be on much longer (centennial to millennial) timescales} \cite{bullock2001recent}.{If such forcings and dampenings are included, then the model will exhibit more complex dynamical behavior (possibly chaotic) instead of sustained finite amplitude oscillations.}

Secondly, we made the simplifying assumption that chemical fluxes remain constant while eddy diffusivity changes in Sec 2.2. This assumption comes from using the steady state chemical model results of Krasnopolsky, which enforce constant fluxes to maintain mass conservation within the model \cite{krasnopolsky1995uniqueness}. As noted earlier in Sec 2.2, there are no fully coupled chemical dynamical models for Venus yet, therefore we are restricted to estimating how dynamical changes affect chemical profiles from such steady state models. In reality, there is no reason for chemical fluxes to remain constant within the dynamically changing atmosphere of Venus, and thus results based on such assumptions will underestimate the magnitude of variability in abundances. 

Thirdly, we have assumed that heating at the cloud-base depends on water abundance alone. While influences from other trace gases maybe small, cloud opacity is another significant source of cloud-base heating change \cite{lee2016sensitivity} and cloud opacity is also known to vary on a timescale of about 150 days \cite{mcgouldrick2007investigation}, comparable to the 120 day timescale estimated for water abundance changes (Eqn. 23). The conditions for oscillations of the nature described here to exist in Venus' atmosphere are based on two tendencies:
\begin{enumerate}
\item A decrease in cloud-base infrared opacity will tend to decrease convective activity
\item Decrease in convective activity will tend to increase cloud-base infrared opacity
\end{enumerate}
Would the changes in cloud mass or thickness yield these tendencies?  If convective strength is weakened, the cloud mass supported by convection should decrease \cite{hashimoto2001predictions}. The reduction in cloud mass leads to cloud droplets falling from the convective middle clouds through the cloud-base (lower cloud) and vaporizing in the sub-cloud region. This represents a net transport of water and sulfuric acid from the middle and lower cloud to the sub-cloud region. Thus, a weakening in convection should result in thinner clouds (lower opacity) and higher sub-cloud water abundances (higher opacity), which has been observed \cite{tsang2010correlations}. Whether these changes result in a net higher or lower cloud-base heating is not easy to predict from first principles arguments and will require coupled microphysics and radiative transfer modeling. The possibility that cloud opacity plays an important role for interannual oscillations definitely exists and should be explored in future studies.

Lastly, within our model we have not considered vertical or horizontal spatial variability, but rather have treated changes in water abundance or eddy diffusivity {that are interpreted as global means}. Nearly all quantities of interest, such as cloud height, water abundance, solar heating and convection show strong dependencies on latitude, longitude, altitude and local time \cite{barstow2012models,encrenaz2019hdo} as well as other transient regional scale changes \cite{tsang2010correlations,arney2014spatially}. Even eddy diffusivity, which we set to be uniform vertically, varies in magnitude depending on whether the region of interest is convective or stable \cite{imamura2001microphysics}, furthermore we have not considered the effects of mixing due to horizontal gradients. Thus, we expect that on Venus the interactions between all these different scales of variability will lead to a rich complexity in oscillations on many different spatial and {temporal} scales as opposed to the single timescale global oscillation we have derived with our simple model. 
\section{Conclusions}

In this paper, {we describe a hitherto unknown atmospheric dynamical mechanism that causes long period variability in the cloud layer of Venus.} We {explored} relationships between cloud-base water abundance and convective strength. {We find that that previously described dependences of infrared radiation on water abundance} \cite{lee2016sensitivity} {and water abundance on eddy mixing} \cite{krasnopolsky2012photochemical} {can be coupled together}. The two resulting prognostic equations represent a recharge-discharge oscillator. {The atmospheric oscillation described here can explain the observed interannual sulfur dioxide variablity at the cloud tops without requiring episodic volcanic injections.} It is a point of interest that a major interannual oscillation on Earth, the El {Ni\~{n}o} Southern Oscillation has also been described using a recharge oscillator paradigm \cite{jin1997equatorial}. The model described in this work bears closer resemblance with the recharge-discharge oscillator describing departures from convective equilibrium (YP12). 

\begin{figure*}
  \includegraphics[width=\columnwidth]{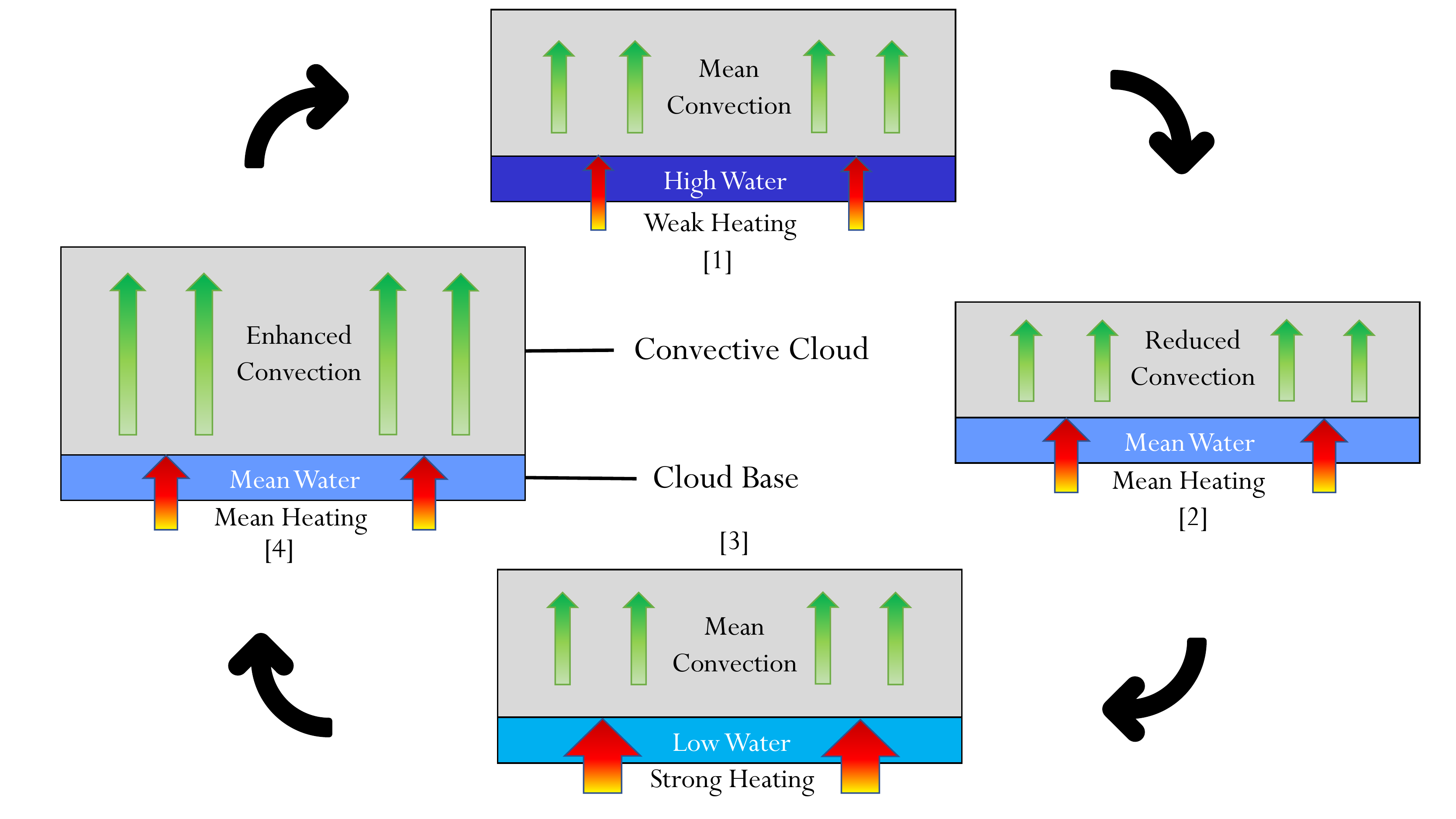}
  \caption{A schematic of the oscillations in the convective layer height and cloud-base water abundance described in this study. The oscillation is sustained because of the finite adjustment times of the cloud-base temperature (and thereby the convective layer height) and the water abundance to the cloud-base heating flux and convective mixing changes respectively.}
  \label{fig:schematic}
\end{figure*}

A schematic of various stages of the oscillation are given in Fig. \ref{fig:schematic}. Physically speaking the oscillation can be described thus: a high water abundance anomaly at the base of the cloud leads to weakening of the cloud base forcing. The {decline} in forcing reduces the convective layer height and vertical mixing in the cloud layer. The {decrease} in vertical mixing reduces the water abundance at the cloud base leading to a low water anomaly, starting the other half of the oscillation. Alternately, the oscillation could also be initiated by a change in convective layer height which then changes the cloud-base water abundance and so forth. In the linear limit (caused by small perturbations to the equilibrium state), the oscillations are simple harmonic. In the non-linear limit the oscillations can be described as a recharge-discharge system. The changes in convective layer height and cloud-base water abundance can strongly affect cloud-top sulfur dioxide concentrations, which have been observed. However, our simplified model cannot quantify such variability, although that is a question of great interest for future studies with coupled chemical and dynamical modeling that builds on the mechanisms proposed in this work.

In the interest of obtaining simple, intuitive relationships, several complex processes were linearized. Thus, the model described here can only be fully justified in the weak perturbation limit where the water and convective strength anomalies are small compared to their equilibrium values. The non-linear limit was also briefly touched upon and shows interesting behavior that maybe closer to the observations, particularly in that the convective strength appears to rise quickly and decrease gradually like the cloud-top sulfur dioxide concentrations \cite{esposito1984sulfur,marcq2013variations}. There is also a weak tendency for the period to depend on the amplitude of the water abundance anomaly, with large anomalies causing oscillations of slightly longer periods. The linear dependence of convective layer height on cloud-base temperature (Eqn. \ref{eqn:heightvar}) is a characteristic feature of many atmospheric shallow water systems \cite[Chap 10 for e.g.,]{heng2017exoplanetary}, and may suggest that this system could be studied under a shallow water framework as a next step. A more thorough treatment of the processes involved with a hierarchy of more sophisticated models would be very useful to better understand such long{-}period oscillations on Venus. 

\acknowledgments
We would like to thank Javier Peralta, Yoshihisa Matsuda and Kevin McGouldrick for comments and discussions which helped improve the paper. PK was funded by the JSPS International Research Fellow program. YJL has received funding from EU Horizon 2020 MSCA-IF No. 841432. \add[editor]{Data and code for reproducing Figures 1-5 are openly available at} \citeA{kopparla2020data}.
\newpage
\bibliography{climaterefs}

\end{document}